\documentclass[prl,aps,twocolumn,10pt,superscriptaddress,showpacs]{revtex4-1}

\usepackage{graphicx}
\usepackage{verbatim}
\usepackage{comment}
\usepackage{amsmath}
\usepackage{amssymb}
\usepackage{stmaryrd}
\usepackage{color}
\usepackage{xfrac}

\begin{document}

\title{Short-Range Correlations and Cooling of Ultracold Fermions\\ in the Honeycomb Lattice}

\author{Baoming Tang}
\affiliation{Department of Physics, Georgetown University, Washington
DC, 20057 USA}
\affiliation{Physics Department, The Pennsylvania State University, 104 Davey Laboratory, 
University Park, Pennsylvania 16802, USA}
\author{Thereza Paiva}
\affiliation{Instituto de Fisica, Universidade Federal do Rio de Janeiro 
Cx.P. 68.528, 21941-972 Rio de Janeiro RJ, Brazil}
\author{Ehsan Khatami}
\affiliation{Department of Physics, Georgetown University, Washington
DC, 20057 USA}
\affiliation{Physics Department, The Pennsylvania State University, 104 Davey Laboratory, 
University Park, Pennsylvania 16802, USA}
\author{Marcos Rigol}
\affiliation{Department of Physics, Georgetown University, Washington
DC, 20057 USA}
\affiliation{Physics Department, The Pennsylvania State University, 104 Davey Laboratory, 
University Park, Pennsylvania 16802, USA}

\begin{abstract}
We use determinantal quantum Monte Carlo simulations and numerical linked-cluster
expansions to study thermodynamic properties and short-range spin correlations
of fermions in the honeycomb lattice. We find that, at half filling
and finite temperatures, nearest-neighbor spin correlations can be stronger in 
this lattice than in the square lattice, even in regimes where the ground state 
in the former is a semimetal or a spin liquid. The honeycomb lattice also exhibits 
a more pronounced anomalous region in the double occupancy that leads to stronger 
adiabatic cooling than in the square lattice. We discuss the implications of these 
findings for optical lattice experiments.
\end{abstract}

\maketitle

In recent years, the isolation of graphene flakes \cite{novoselov_geim_04} has
generated a revolution in solid state physics \cite{castroneto_guinea_review_09}. Graphene is 
an atom thick structure with carbon atoms arranged in a honeycomb lattice
geometry, which features low energy excitations that are massless Dirac fermions. Given 
its reduced coordination number, graphene has also opened a new venue to create exotic 
quantum phases. Based on quantum Monte Carlo (QMC) simulations of the half filled
one-band Hubbard model, Meng {\it et al.}~\cite{meng_lang_10} have argued that a spin-liquid 
ground state may be realized in this lattice geometry at intermediate interaction strengths. 
Earlier works had found the ground state to be a semimetal in the weakly-interacting regime 
and a Mott insulator with long-range antiferromagnetic (AF) correlations in the strongly-interacting 
regime \cite{sorella_tosatti_92,paiva_scalettar_05}. The finding of an intermediate 
spin-liquid phase, recently challenged by another QMC study that considered larger lattice 
sizes~\cite{s_sorella_12}, motivated much theoretical work in related models
\cite{mulder_ganesh_10,ganesh_sheng_11,cabra_lamas_11,mosadeq_shahbazi_11,
albuquerque_schwandt_11,clark_abanin_11,f_mezzacapo_12,reuther_abanin_11,w_wu_12}.

Experiments on fully tunable artificial graphene-like lattices now offer a 
pathway to study the physics above, and more, in a controlled 
way~\cite{m_gibertini_09,a_singha_11,tarruel_greif_12,gomes_mar_12}. 
Motivated by those experimental advances, especially by the
availability of an ultracold lattice fermion setup \cite{tarruel_greif_12}, where
on-site interactions, hopping amplitudes, doping, and temperature can be fully controlled 
using Feshbach resonances, changing the lattice depth and the number of fermions in the gas, 
and varying the cooling time \cite{bloch_dalibard_review_08}, respectively, we study thermodynamic 
properties and short-range correlations of two-component correlated fermions in the honeycomb lattice. 

We show that such a system exhibits several unexpected properties when compared with its square 
lattice counterpart. For example, at a finite temperature, it may be less compressible in the 
weakly interacting regime where its ground state is a semimetal, while the latter becomes less 
compressible in the presence of strong interactions when both lattices have an insulating ground 
state. We also identify temperature regimes in which, surprisingly, 
(i) nearest-neighbor (NN) spin correlations are stronger and (ii) a more significant anomalous region 
can be seen in the derivative of the double occupancy with respect to temperature, in the honeycomb 
lattice than in the square lattice.

We consider the one-band Hubbard Hamiltonian
\begin{equation}
 \hat{H}=-t\sum_{\langle i,j \rangle \sigma}(\hat{c}^\dagger_{i\sigma}\hat{c}_{j\sigma}+ \text{H.c.})
         +U\sum_{i}\hat{n}_{i\uparrow}\hat{n}_{i\downarrow}\, ,\label{ham}
\end{equation}
where standard notation has been used \cite{imada_fujimori_98}. At half filling, in the square lattice, 
the ground state of this model is an AF Mott insulator for any $U>0$ \cite{imada_fujimori_98}, while, 
in the honeycomb lattice, it has been recently argued to be a semimetal for $0\leq U/t\lesssim 3.5$,
an AF Mott insulator for $U/t\gtrsim 4.3$, and a gapped spin liquid in between \cite{meng_lang_10}.

In this Letter, to study the properties of the Hamiltonian [Eq.\eqref{ham}] in the honeycomb and 
square lattices, we utilize two unbiased computational approaches, the determinantal quantum Monte Carlo 
(DQMC) technique \cite{scalapino_sugar_81a,loh_gubernatis_92,muramatsu_99} and numerical 
linked-cluster expansions (NLCEs) \cite{rigol_bryant_06_25,rigol_bryant_07_30,rigol_bryant_07_31}.
DQMC simulations are performed in finite-size systems (with 100 and 96 sites 
for the square and honeycomb lattices, respectively) using a small discretized imaginary time 
($\Delta \tau \times t=0.05$). NLCE calculations, on the other hand, 
provide exact results in the thermodynamic limit but converge down to a temperature 
that is determined by the divergence of correlations and the largest cluster sizes
that we can consider. Here, we include clusters up to the ninth order in the site expansion 
and use Wynn and Euler resummation algorithms to extend the region of convergence to lower 
temperatures \cite{rigol_bryant_06_25,rigol_bryant_07_30,pedog}. DQMC calculations and NLCEs are complementary 
as the former provides
more accurate results down to lower $T$ for $U\lesssim w$, where $w$ is the noninteracting 
bandwidth ($w=6t$ for the honeycomb
lattice and $w=8t$ for the square lattice) while the latter is better 
suited for $U>w$ \cite{khatami_rigol_11_63}. In the region where DQMC statistical errors are small 
and NLCEs converge, we obtain an excellent agreement between both approaches.

\begin{figure}[!t]
\includegraphics[width=0.4\textwidth]{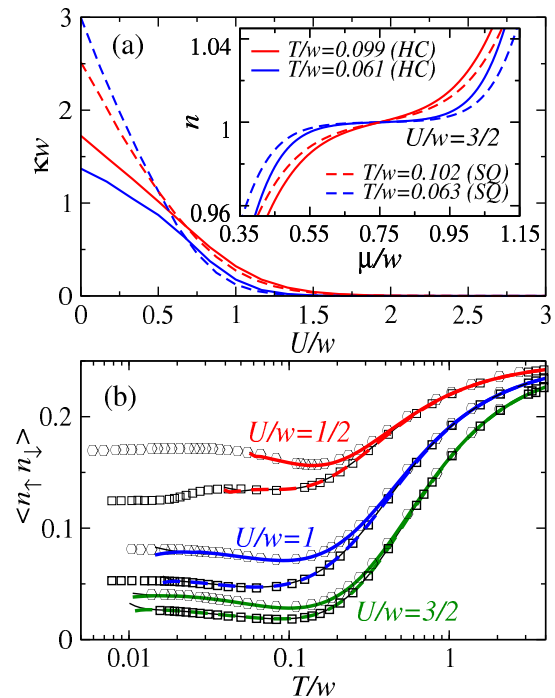}
\caption{(Color online) (a) NLCE results for the compressibility vs $U$ in the honeycomb (HC) and 
square (SQ) lattices, at half filling, for two values of $T/w$ that are very close in both lattices. 
{\em Inset-} Equation of state for $U/w=3/2$ and the same two values of $T/w$ as in the main 
panel. These results were obtained after three cycles of improvement of Wynn's 
resummation algorithm \cite{rigol_bryant_07_30}. The zero chemical potential, which corresponds 
to half filling for the particle-hole symmetric Hamiltonian, is shifted by $U/2$ for the 
nonsymmetric representation of the Hamiltonian in Eq.~\eqref{ham}. (b) DQMC (symbols) and NLCE (lines) results for  
$\langle\hat{n}_{\uparrow}\hat{n}_{\downarrow}\rangle$ vs $T$ in both lattices at half filling for 
$U/w=1/2$, $1$, and $3/2$. 
Hexagons (Squares) and solid (dashed) lines correspond to honeycomb (square) lattice. Statistical error 
bars for DQMC data are shown only when they are greater than the symbol size. The NLCE results were 
obtained using Euler resummation, and we report the last order (thick lines) and the one to last 
order (black thin lines). }
\label{fig:Dob}
\end{figure}

In optical lattice experiments, single site addressability 
\cite{bakr_peng_10,sherson_weitenberg_10} makes possible an accurate determination of the 
equation of state [density ($n$) vs chemical potential ($\mu$)] of lattice Hamiltonians 
of interest. This equation of state determines the shape of the experimental density 
profiles and, when obtained at low enough temperatures, 
allows one to identify the presence of a single particle gap in the spectrum. In the inset of 
Fig.~\ref{fig:Dob}(a), we show the equation of state in the square and honeycomb lattices for 
$U/w=3/2$, which is beyond the critical value for the formation of the Mott insulator in the latter, 
and for two values of $T/w$ that are very close in both lattices. With decreasing temperature, 
$n$ vs $\mu$ reveals the single-particle gap in the Mott phase by exhibiting a region 
in which $n$ barely changes when changing $\mu$. As expected from their phase diagrams, 
that gap is greater in the square lattice than in the honeycomb lattice. This results in 
the former system being less compressible than the latter at half filling and finite $T$
for large values of $U$. 

By decreasing $T$ for small $U$, the compressibility 
($\kappa=\partial n/\partial\mu$) also reveals the vanishing of the density of states in the 
semimetallic phase. This is shown in the main panel of Fig.~\ref{fig:Dob}(a), where, for weak 
interactions, the compressibility in the honeycomb lattice is seen to decrease with decreasing 
temperature ($\kappa\rightarrow0$ as $T\rightarrow 0$). This behavior is to be contrasted with 
the one in the square lattice, where $\kappa$ increases as $U\rightarrow 0$ and $T\rightarrow 0$, 
signaling the metal insulator transition \cite{imada_fujimori_98}. Note that, for finite $T$,
the behavior above leads to a region in $U$ where $\kappa$ is smaller in the honeycomb 
lattice than in the square lattice, despite the fact that in such a region the ground 
state in the former may be a semimetal while in the latter is an insulator. This 
can be understood given the difference between dispersion relations in the two systems 
which, at low $T$, can lead to less states being available in the honeycomb lattice
than in the square lattice.

Another quantity of much interest, which can also be measured in experiments with ultracold fermions 
\cite{jordens_strohmaier_08}, is the double occupancy $\langle\hat{n}_{\uparrow}\hat{n}_{\downarrow}\rangle$. 
At half filling, $\langle\hat{n}_{\uparrow}\hat{n}_{\downarrow}\rangle$ is expected to decrease with
decreasing temperature. This can be seen in Fig.~\ref{fig:Dob}(b), where we plot DQMC (symbols) 
and NLCE (lines) results for the double occupancy vs $T$ for three values of $U$ in the honeycomb
and square lattices. (Note the excellent agreement between the results obtained utilizing the two approaches.) 
At high temperatures, $\langle\hat{n}_{\uparrow}\hat{n}_{\downarrow}\rangle$ is essentially the same 
for both geometries. However, as the double occupancy decreases when reducing $T$, one can see that 
the results in the honeycomb lattice depart from, and remain at higher values than, those 
in the square lattice. As this occurs, an upturn can be seen in the double occupancy with decreasing 
$T$. Especially for small $U/w$, this upturn is more pronounced in the honeycomb lattice than in the 
square lattice (note that for $U/w=1/2$, it is absent in the latter geometry). 

\begin{figure}
\includegraphics[width=0.45\textwidth]{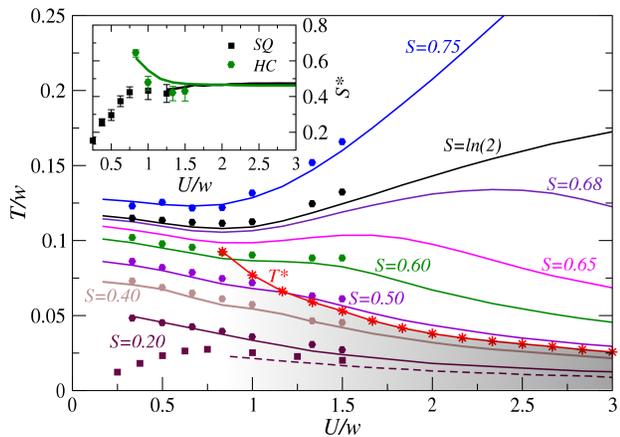}
\caption{(Color online) Isentropic curves for the temperature vs $U/w$ in the honeycomb 
lattice at half filling at several constant entropies. The crossover scale, $T^*$, from NLCE is only 
depicted in the regime where 
the ground state is a Mott insulator with long-range AF correlations.
For $S$=0.20, we also show results for the square lattice (dashed line and 
square symbols). 
The inset shows the entropy per particle at $T=T^*$ vs $U/w$. Lines (Symbols) correspond to NLCE 
(DQMC) results.}
\label{fig:T_vs_U}
\end{figure}

The existence of a region in temperature in which there is an anomalous 
$d\langle\hat{n}_{\uparrow}\hat{n}_{\downarrow}\rangle/dT<0$ has been discussed in the context 
of the Hubbard model in the square lattice. Early dynamical mean-field theory calculations 
identified a significant anomalous region \cite{werner_parcollet_95}, which was later found to 
be marginal in DQMC \cite{paiva_scalettar_10} and NLCE \cite{khatami_rigol_11_63} 
calculations in two dimensions (2D). Interest in the existence of such a region developed as it 
signals adiabatic cooling with increasing $U$. This follows from the relation 
$\partial S/\partial U=-\partial \langle\hat{n}_{\uparrow}\hat{n}_{\downarrow}\rangle/\partial T$ 
\cite{werner_parcollet_95}, which implies that at constant $T$, the entropy ($S$) increases 
(or, that at constant $S$, the temperature decreases) with increasing $U$. 
DQMC \cite{paiva_scalettar_10} and NLCE \cite{khatami_rigol_11_63} 
calculations have also shown that, starting with short-range spin correlation for small 
values of $U$, one can generate exponentially large AF correlations by adiabatically increasing 
$U$, despite the fact that there is almost no cooling for weak interactions. However, 
the entropy per particle needs to be $S\lesssim 0.5$.

We plot in Fig.~\ref{fig:T_vs_U}, the isentropic
curves in the $T-U$ plane for the honeycomb lattice. By comparing those results with the ones for the 
square lattice (see Refs.~\cite{paiva_scalettar_10,khatami_rigol_11_63} and the results
for $S=0.2$ in Fig.~\ref{fig:T_vs_U}), it becomes apparent 
that adiabatic cooling is more significant in the honeycomb lattice for small values of $U$.
This occurs in the absence of a Mott insulating ground state and where 
the available number of states at any given $T$ in the honeycomb lattice is smaller 
than in the square lattice [see the compressibilities in Fig.~\ref{fig:Dob}(a)].
One may wonder if this could ease the realization of exponentially large AF correlations 
in the honeycomb lattice in comparison to the square lattice, where it remains a major 
experimental goal \cite{esslinger_review_10}. The region with exponentially 
large correlations can be identified 
from $T^*$, which is the temperature at which the uniform susceptibility is maximal for $U$ 
beyond the critical value for the formation of the Mott insulator. $T^*$ is also plotted in 
Fig.~\ref{fig:T_vs_U} and shows that an entropy per particle $S\lesssim 0.6$ is needed 
to generate exponentially large correlations in the honeycomb lattice. This is close to, 
but above, the entropy required in the square lattice. The entropy per particle 
at $T^*$ in the square and honeycomb lattices for half filled systems, $S^*$, is shown in the inset 
of Fig.~\ref{fig:T_vs_U}. Beyond $U/w=1$, $S^*$ can be seen to be almost the same 
in both lattice geometries ($S^*\lesssim 0.5$).

\begin{figure}[!t]
\includegraphics[width=0.46\textwidth]{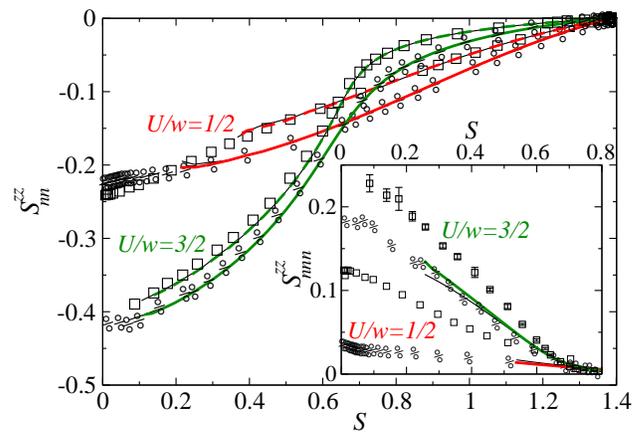}
\caption{(Color online) (main panel) Nearest-neighbor spin correlations and (inset) 
next-nearest-neighbor spin correlations in the honeycomb and square lattices at 
half filling as a function of entropy, for $U=w/2$ and $3w/2$. Lines and symbols 
are the same as in Fig.~\ref{fig:Dob}(b). 
Note that, for $S^{zz}_{nnn}$ in the square lattice, only QMC results are shown.}
\label{fig:Szz}
\end{figure}

Probing long-range AF correlations turns out to be very challenging in optical 
lattice experiments. As a first step towards this goal, and towards identifying the 
AF Mott insulator in the Hubbard model on the square lattice, experiments have 
already measured NN spin correlations $S^{zz}_{nn}$
\cite{trotzky_chen_10,greif_tarruell_11}. They increase as the temperature is lowered and 
can be significant even before long-range order sets in the system. In Fig.~\ref{fig:Szz}, 
we plot $S^{zz}_{nn}$ in the square and honeycomb lattices vs $S$ for two different values of
$U/w$, one below and one above the critical value for the formation of the Mott insulator 
in the honeycomb lattice. That figure shows that, unexpectedly, there is an extended region
in entropies where $|S^{zz}_{nn}|$ are greater in the honeycomb lattice 
than in the square lattice, and that this happens even when the ground state in the former 
is a semimetal or a spin liquid while it is an AF Mott insulator in the latter. 
At very low entropies, we find that ,ultimately, $|S^{zz}_{nn}|$ in the square lattice becomes 
greater than in the honeycomb lattice, but the entropy at which this occurs becomes smaller 
as $U$ increases.

Our results imply that strong NN spin correlations can be more easily observed in experiments 
in the honeycomb lattice than in the square lattice. They also make evident that an enhancement 
of $|S^{zz}_{nn}|$ should not be taken as a signature of the Neel state, which does not exist in 
the honeycomb lattice for $U/w<0.72$, where $|S^{zz}_{nn}|$ is greater than in the square lattice 
(for entropies per particle that are currently achievable experimentally). This is because such 
an enhancement can be a very local effect. We have also calculated next-nearest-neighbor 
correlations, $S^{zz}_{nnn}$, in both lattices (see the inset of Fig.~\ref{fig:Szz}) and 
found them to be always stronger in the square lattice than in the honeycomb lattice.

Cooling fermions in optical lattices to realize the Neel state is currently one of the 
main experimental challenges~\cite{esslinger_review_10}. To that purpose, one can take advantage 
of the fact that the system is inhomogeneous [a term $\sum_{i\sigma}Vr_{i}^{2}~\hat{n}_{i\sigma}$, 
where $V$ is the strength of the trapping potential and $r_{i}$ is the distance of each lattice 
site to the center of the trap, needs to be added to Eq.~\eqref{ham}] and that this implies 
that the entropy is unevenly distributed in the gas \cite{esslinger_review_10}. Based on that idea, 
two recent works, one on the square lattice \cite{khatami_rigol_11_63} and the other on the cubic 
lattice \cite{paiva_loh_11}, have shown that starting with a system with high density in the center 
of the trap ($n\sim 2$) and with an average entropy per particle larger than $S^*$, one can achieve 
a Mott insulator in the center of the trap with a local entropy smaller than or equal to $S^*$ by 
adiabatically decreasing the confining potential.
The excess entropy is then stored in the compressible domains with $n<1$.

\begin{figure}[!t]
\includegraphics[width=0.48\textwidth]{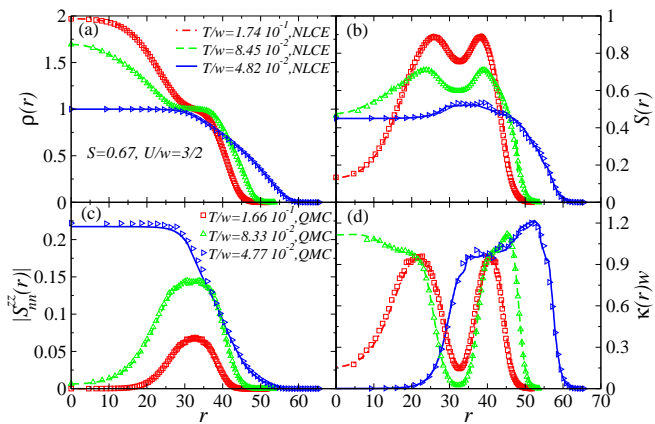}
\caption{(Color online)  (a) Density, (b) entropy, (c) NN spin correlations, and (d) compressibility 
for trapped honeycomb lattice systems with $N=6.7\times10^{3}$ particles, an average entropy per particle $S=0.67$, $U=3w/2$, and 
$V/w=1.5\times10^{-3},1.0\times10^{-3}$, and $3.8\times10^{-4}$ (in order of decreasing temperature). 
DQMC results are depicted as symbols and NLCE results as lines. Note that, due to the finite 
$T$ grids used in DQMC and NLCE calculations, the values of $T$ in both approaches are very close 
but not identical.}
\label{fig:trapped}
\end{figure}

In Fig.~\ref{fig:trapped}, we use the local density approximation (LDA), combined with DQMC and 
NLCE calculations of the homogeneous system, to show how the cooling mechanism discussed above works
in the honeycomb lattice. (For temperatures like the ones studied here,
DQMC calculations have shown that LDA is a good approximation on the square lattice 
\cite{chiesa_varney_11_54}.) Figure \ref{fig:trapped} depicts the evolution of the local density (a),
local entropy (b), NN spin correlations (c), and the local compressibility (d) as one 
reduces the trapping potential adiabatically in a system with $U/w=3/2$ and in which the average entropy 
per particle is $S=0.67$. This entropy per particle is higher than $S^*=0.47$ for $U/w=3/2$.
One can see that, as $V$ is reduced, the density in the center of the trap 
changes from nearly that of a band insulator to that of a Mott insulator [Fig.~\ref{fig:trapped}(a)]. 
At the same time, the entropy in the Mott insulating region becomes of the order of, or smaller than, $S^*$, 
with the excess entropy being moved to the metallic wings [Fig.~\ref{fig:trapped}(b)]. This results 
in strong NN correlations in the Mott insulating domain [Fig.~\ref{fig:trapped}(c)]
and a vanishing compressibility in the same region [Fig.~\ref{fig:trapped}(d)]. Our results for 
a specific trapping potential and number of particles (similar to the ones in current experiments) 
can be extended to other values of the trapping potential and number of particles through 
the use of the characteristic density 
\cite{rigol_muramatsu_03_5,rigol_muramatsu_04_11}.

In summary, we have used DQMC and NLCEs to study experimentally relevant 
thermodynamic properties and spin correlations of the Hubbard model in the honeycomb lattice. We 
find that, at half filling and weak interactions, the compressibility in this lattice may be 
smaller than in the square lattice at low 
$T$, despite the fact that the ground state in the 
former is a semimetal and in the latter an insulator. We also find that the honeycomb lattice
exhibits a more significant anomalous region with $d\langle\hat{n}_{\uparrow}\hat{n}_{\downarrow}\rangle/dT<0$
than the square lattice, which leads to a stronger adiabatic cooling in the former lattice 
geometry. Remarkably, NN spin correlations in the honeycomb lattice are stronger 
than in the square lattice in an extended region of entropies for all 
$U$. We discussed how these findings are reflected in optical lattice experiments.

\begin{acknowledgments}
This work was supported by NSF Grant No.~OCI-0904597 (BT, EK, and MR),
and by CNPq, FAPERJ and INCT on Quantum Information (TP). TP thanks 
R. T. Scalettar and L. G. Marcassa for discussions.
\end{acknowledgments}

\end{document}